\documentclass[apjl]{emulateapj}
\slugcomment{Accepted for publication in Astrophysical Journal Letters}
\shorttitle{THE ARCHES CLUSTER MASS FUNCTION}
\shortauthors{KIM ET AL.}

\def\spose#1{\hbox to 0pt{#1\hss}}
\newcommand\lsim{\mathrel{\spose{\lower 3.0pt\hbox{$\mathchar"218$}}
     \raise 2.0pt\hbox{$\mathchar"13C$}}}
\newcommand\gsim{\mathrel{\spose{\lower 3.0pt\hbox{$\mathchar"218$}}
     \raise 2.0pt\hbox{$\mathchar"13E$}}}
\newcommand\msun{{\rm \,M_\odot}}

\begin{document}
\title{THE ARCHES CLUSTER MASS FUNCTION}
\author{Sungsoo S. Kim,\altaffilmark{1} Donald F. Figer,\altaffilmark{2}
Rolf P. Kudritzki,\altaffilmark{3} and F. Najarro\altaffilmark{4}}
\altaffiltext{1}{Dept. of Astronomy \& Space Science, Kyung Hee University,
Kyungki-do 446-701, Korea; sungsoo.kim@khu.ac.kr}
\altaffiltext{2}{Chester F. Carlson Center for Imaging Science,
Rochester Institute of Technology, 54 Lomb Memorial Drive, Rochester,
NY 14623-5604}
\altaffiltext{3}{Institute for Astronomy, University of Hawaii,
2680 Woodlawn Drive, Honolulu, HI 96822}
\altaffiltext{4}{Instituto de Estructura de la Materia, Consejo Superior
de Investigaciones Cientificas, Calle Serrano 121, 28006 Madrid, Spain}

\begin{abstract}
We have analyzed $H$ and $K_s$--band images of the Arches cluster obtained
using the NIRC2 instrument on Keck with the laser guide star adaptive optics
(LGS AO) system.  With the help of the LGS AO system, we were able
to obtain the deepest ever photometry for this cluster and its neighborhood,
and derive the background-subtracted present-day mass function (PDMF) down to
$1.3 \msun$ for the $5 \arcsec$--$9\arcsec$ annulus of the cluster.  We find
that the previously reported turnover at $6 \msun$ is simply due to a local
bump in the mass function (MF), and that the MF continues to increase down
to our 50~\% completeness limit ($1.3 \msun$) with a power-law exponent of
$\Gamma = -0.91$ for the mass range of $1.3 < M/\msun < 50$.  Our numerical
calculations for the evolution of the Arches cluster show that the $\Gamma$
values for our annulus increase by 0.1--0.2 during the lifetime of the cluster,
and thus suggest that the Arches cluster initially had $\Gamma$ of
$-1.0 \sim -1.1$, which is only slightly shallower than the Salpeter value.
\end{abstract}
\keywords{Galaxy: center --- open clusters and stellar associations: general
--- stars: luminosity function, mass function --- celestial mechanics, stellar
dynamics}

\section{INTRODUCTION}
\label{sec:introduction}

The stellar initial mass function (IMF) is the most primary product of the
star formation process.  With this one relation, it is possible to directly
probe the crucial predictions of star formation models.  Surprisingly,
the IMF is approximately universal, following the Salpeter law (Salpeter 1995)
for masses between 1 and $120 \msun$ (Kroupa 2002).  On the low mass
end, there appears to be a universal rollover around $0.8 \msun$, and the
high mass end seems to be truncated by a sharp cutoff near $150 \msun$
(Weidner \& Kroupa 2004; Figer 2005).

However, star formation theories predict that the lower mass cutoff ($m_l$)
should be a function of environmental parameters, i.e. magnetic field strength
and cloud temperature (e.g., Bonnell, Larson, \& Zinnecker 2006).
In the extreme environment of the Galactic center (GC), some of these
models predict an elevated lower mass cutoff/rollover with respect to
that observed in the disk (Morris 1993).  If this prediction is true,
then the GC should have an abnormally bright lower mass cutoff, something
that has not been {\it directly} observed anywhere in the Universe.

The best places to trace the IMF near the GC are the two young star clusters
therein: the Arches and Quintuplet clusters (see Figer et al. 1999 for
findings and early studies on these clusters).  These clusters are very young
(2--4~Myr), compact ($\lsim 1$~pc), and only 20--30~parsecs away from the GC
in projection, while they appear to be as massive as the smallest Galactic
globular clusters ($\sim 2 \times 10^4 \msun$; Kim et al. 2000).  Of the
two clusters, the Arches is preferred for studies of the low-end mass
function (MF) as the Quintuplet is significantly more dispersed.

While the Arches cluster's compactness makes its stars easily identifiable
for their high spatial density projected onto the background population, this
same property makes confusion a problem at faint magnitudes. Note for scale
that a $1 \msun$ dwarf has ${\rm K} \simeq 21$  in the cluster.  For instance,
the luminosity function (LF) obtained from {\it Hubble Space Telescope}
({\it HST}) NICMOS observations by Figer et al. (1999) for the whole cluster
was 50~\% complete at $6 \msun$.  The adaptive optics (AO) science
demonstration data observed with the Gemini/Hokupa'a system were analyzed by
Yang et al. (2002) and Stolte et al. (2002), who obtained a 50~\% completeness
level at $5.6 \msun$ and a 75~\% completeness level at $6 \msun$ for the whole
cluster, respectively.  Stolte et al. (2005) have analyzed the natural
guide star (NGS) AO observations of the cluster with the Very Large
Telescope NAOS/CONICA system, and derived the MF of the intermediate
region of the cluster ($5 \arcsec$--$10 \arcsec$) down to $2.3 \msun$ at
a 50~\% completeness limit.

Although the last study reached the faintest magnitude to date, due to
the lack of suitable NGS candidates for neighboring fields, it lacks the
observations of nearby background populations, which should be subtracted
from the LF/MF of the cluster field to obtain the true cluster LF/MF.
It had no choice but to compare the MF from the central and intermediate
regions of the cluster ($r<10\arcsec$) with that from the outer region
($r>13.4\arcsec$) to infer the MF of the cluster only.  However, while
there are almost no bright cluster stars in this outer region, a non-negligible
number of faint cluster stars may be present there due to mass
segregation.  For example, the comparison between {\it HST} observations
and $N$-body simulations by Kim et al. (2000) indicates that the tidal
radius of the Arches cluster is $25 \arcsec$--$30 \arcsec$.  Thus a reliable
estimation of the LF/MF of the nearby background populations would require
observations of control fields that are at least $30 \arcsec$ away from
the cluster center.

For this reason, we have carried out Keck/NIRC2 laser guide star (LGS)
AO observations of the Arches cluster and its nearby background
populations to take advantage of the LGS AO system that does not have
the distance limit from NGS candidates.  Here, we present the analysis
of these observations along with results from several relevant $N$-body
simulations.

\section{OBSERVATIONS AND DATA REDUCTION}
\label{sec:observations}

\begin{figure}
\centering
\resizebox{7.0cm}{!}{\includegraphics{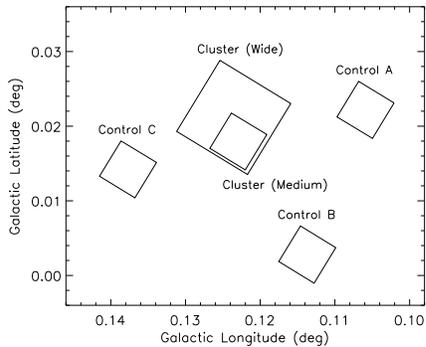}}
\caption{\label{fig:location}
Locations of our images in the galactic coordinates.  The control fields,
which were taken to estimate the stellar density of the nearby background
population, are separated $\sim 60 \arcsec$ from the cluster center.}
\end{figure}

The images were obtained using the Keck/NIRC2 LGS AO system on 2006 May
4 \& 5.  The Arches cluster ($\alpha={\rm 17^h 45^m 50^s .59}$, $\delta={\rm
-28\arcdeg 49'20''.3}$; J2000) was observed with the medium ($20 \arcsec
\times 20 \arcsec$, 0\arcsec .0198 pixel$^{-1}$) and wide ($40 \arcsec \times
40 \arcsec$, 0\arcsec .0397 pixel$^{-1}$) field cameras, and three nearby
control fields, separated $\sim 60 \arcsec$
from the cluster center, were imaged with the medium field camera in order to
sample the background population.  The locations of these fields in galactic
coordinates are shown in Figure~\ref{fig:location}.  All fields were imaged
in $H$ and $K_s$ bands in a 9-point dither pattern with a leg size of
$2 \arcsec$.
The multiple correlated double sampling (MCDS) mode was used with 10 coadds
and a 10~s integration time per coadd, giving a total exposure time of 900~s
per field.  Obtained Strehl ratios for the $K_s$ ($H$) images are 0.17 (0.08),
0.19 (0.08), 0.25 (0.06), 0.13 (0.06), and 0.12 (0.06) for the medium cluster
field, control fields A, B, C, and wide cluster field, respectively,
resulting in FWHMs of 70--100~mas for $K_s$, and 61--98~mas for $H$.

For data reduction, the dithered images were first combined by maximizing
the cross-correlation.  Star finding, point-spread function (PSF) building,
and PSF fitting procedures were performed using the DAOPHOT package within
the Image Reduction and Analysis Facility (IRAF).  The medium field camera PSFs
were moderately elongated toward the center of the image due to a relatively
small isoplanatic patch obtained during the observations, thus we have
allowed the PSFs to be quadratically variable along both axes of the field.

Typical DAOPHOT magnitude errors were less than 0.05~mag down to $K_s \simeq
19$ for stars in the cluster field with $r>5 \arcsec$ ($K_s \simeq 18$ for the
wide field camera) and $K_s \simeq 19.5$ for stars in the nearby fields.
We do not include stars with errors greater than 0.2~mag ($K_s \gsim 21.5$)
in our analyses.  The elongation of the PSF of the medium camera was
significant at the four corners of the images, so we excluded stars farther
than $10 \arcsec$ from the image center.

Photometric calibration was done against the {\it HST}/NICMOS photometry
by Figer et al. (1999) assuming that $H$ and $K_s$ magnitudes are identical
to NICMOS F160W and F205W magnitudes for the stars in our fields (Kim et al.
2005 and Stolte et al. 2002 indicate that the transformation from
$H$ and $K'$ to $F160W$ and $F205W$ filters has a very weak dependence on
color for the stars in the Arches cluster).

Completeness tests were performed for all images by adding artificial
stars to the observed images with the PSF obtained from our reduction.
800 artificial stars were used per magnitude bin (a total of 7200 stars),
but only 200 stars
were added per test image.  For our analyses in the following sections,
we will only consider magnitude and mass bins with a completeness fraction
larger than 50~\%.  Due to the relatively irregular shape of the
AO PSFs, the DAOPHOT star-finding algorithm resulted in many bogus stars
around bright stars, and we have excluded from our photometry the faint stars
that are $\sim 4$~mag fainter than a nearby (within $\sim 0.55''$ for
the medium field camera and $\sim 0.8''$ for the wide field camera),
brighter star.  Since the central
region of the cluster is dominated by bright stars, we will analyze the LF/MF
of the intermediate region ($5 \arcsec$--$9 \arcsec$) where the photometry
of the faint stars is limited by the background confusion.  But there are
still a few bright stars in the intermediate region that can hamper the
recovery of nearby fainter stars.  Since we are more interested in the
incompleteness of faint stars by background confusion than that by nearby
bright stars, we exclude artificial stars that are $\sim 4$~mag fainter and
within $\sim 0.55''$ ($\sim 0.8''$ for the wide field) from a nearby star
from calculating the completeness fractions.

\begin{figure}
\centering
\resizebox{7.6cm}{!}{\includegraphics{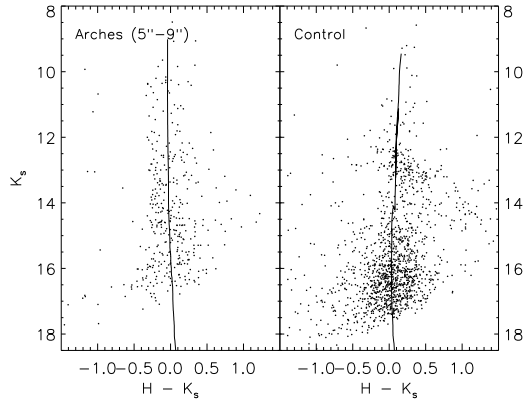}}
\caption{\label{fig:cm}
Dereddened color-magnitude diagram of the stars with DAOPHOT magnitude errors
less than 0.2 mag in the $5 \arcsec$--$9 \arcsec$ annulus of the cluster
field ({\it left}) and in control fields A, B, and C.
Also shown are the solar-metallicity isochrones for 2~Myr ({\it left})
and 1~Gyr ({\it right}) from the Geneva set of models.  The limiting magnitude
in $H$ is $\sim 1$~mag brighter than $K_s$, so the number of stars shown
in the left panel is much smaller than that used for deriving our mass
function later, which is based on the $K_s$-band photometry only.}
\end{figure}

\section{RESULTS}
\label{sec:results}

\subsection{Extinction}

For our analyses, we adopt the extinction law by Rieke, Rieke, \& Paul (1989;
$A_H=1.56 A_K$) and the distance modulus for the GC of 14.52~mag (8~kpc).
Stolte et al. (2002) reported color variation of stars as a function of
the distance from the cluster center and interpreted it as a local
depletion of dust (and thus a smaller extinction) at the cluster center
due to winds from massive stars or photo-evaporation of dust grains
by the intense UV radiation field.  Our data confirm such variation, and
we determine the extinction values for the $5 \arcsec$--$9 \arcsec$ annulus
of the cluster
field, the outer ($r > 15 \arcsec$) region of the cluster field, and three
control fields separately.  Then we assume that stars in the same field
or annulus have the same extinction value.

We compare the 2~Myr, solar-metallicity isochrone from the Geneva set of
models (Lejeune \& Schaerer 2001) to our color-magnitude ($H-K_s$ vs. $K_s$)
diagram (Fig.~\ref{fig:cm}) and obtain an average extinction at $K_s$ of
$3.1 \pm 0.19$~mag for
the $5 \arcsec$--$9 \arcsec$ annulus of the cluster field, which is consistent
with the extinctin value estimated by Stolte et al. (2005) for the same region.

Our control fields and outer region of the cluster field are expected
to be composed of stars with a variety of ages and metallicities (Figer
et al. 2004).  However, $H-K_s$ colors of the stars with intrinsic $K_s$
magnitudes of 1--3~mag, which have undereddened, apparent $K_s$ magnitudes
of 18--20~mag at the GC, vary only little ($\lsim 0.05$~mag; see Kim et al.
2005, for example) for various ages and metallicities.  So we compare the
1~Gyr, solar-metallicity isochrone to the $H-K_s$ color distribution 
of stars with underredened, apparent $K_s$ of 18--20~mag, and find that
our control fields A, B, \& C have $A_{K_s}$ of $2.9 \pm 0.22$, $3.1 \pm 0.17$,
\& $2.5 \pm 0.25$~mag, respectively, and the outer region of the cluster
field has $A_{K_s}=3.0 \pm 0.25$~mag (quoted uncertainties are the
averages of the extinction spread).  Since the amounts of extinction spread
are similar in our cluster and control fields, the differential extinction
in each field will not cause a problem to our subtraction of background
population from the cluster field.

\begin{figure}
\centering
\resizebox{7.6cm}{!}{\includegraphics{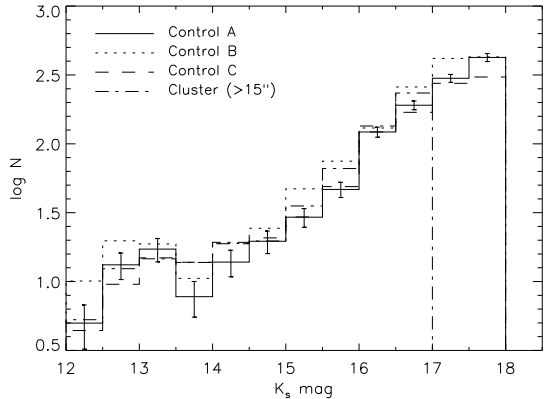}}
\caption{\label{fig:lfkc}
Dereddened, completeness corrected $K_s$--band luminosity function for
control fields A, B, and C, and the outer region ($r>15 \arcsec$) of the cluster
field from the wide field camera, scaled to the area of $5 \arcsec$--$9
\arcsec$ annulus.  The histograms are shown only for magnitude bins whose
recovery fraction is more than 50~\%.  Also shown are the Poisson error bars
for the control field A.}
\end{figure}

\subsection{Background Population}

Figure~\ref{fig:lfkc} compares the dereddened, completeness-corrected
$K_s$--band LFs for the control fields and that for the outer region
($r>15\arcsec$) of the cluster field obtained with the wide field
camera.  While the LFs of control fields A and C are quite similar,
control field B and the outer region of the cluster field generally have
higher LFs than the former.  This is probably because the control field B
is located closer to the Galactic plane where the stellar density is larger
and a part of the outskirt and tidal tail of the Arches cluster might be
included in our outer cluster field.  We take an average of control fields
A, B, and C as the LF of the background population of the cluster. Note that
this will probably overestimate the background population toward the
Arches cluster slightly since we do not have a control field that is located
at the opposite side of the control field B, and thus our estimate
for the cluster LF/MF would be a lower limit.

\begin{figure}
\centering
\resizebox{7.6cm}{!}{\includegraphics{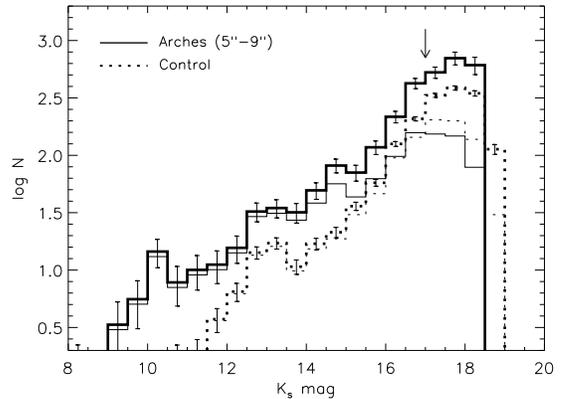}}
\caption{\label{fig:lf}
Dereddened $K_s$-band luminosity functions
for the $5 \arcsec$--$9 \arcsec$ annulus of the cluster field ({\it solid})
and the average of the control fields A, B, and C ({\it dotted}) scaled for
the area of $5 \arcsec$--$9 \arcsec$ annulus.  {\it Thin lines} are for
the raw LFs and {\it thick lines} are for the completeness-corrected LFs.
The Poisson error bars are shown for the latter.  The arrow indicates the 50~\%
completeness limit for the $5 \arcsec$--$9 \arcsec$ annulus of the cluster
field.}
\end{figure}

Figure~\ref{fig:lf} compares the dereddened $K_s$--band LFs for the
background population (an average of control fields A, B, and C) and
that for the intermediate region of the cluster field
obtained with the medium field camera.  The figure shows both raw and
completeness-corrected LFs, and the latter of the cluster field is clearly
larger than that of the background population at least down to $K_s=17$~mag,
which is the 50~\% completeness limit for the cluster field.  The $H$--band LFs
give a similar result except that their 50~\% completeness
limit is 0.5--1 mag brighter than for the $K_s$ band (the limit in $H$ itself
is 1~mag brighter than that in $K_s$, but the typical apparent $H-K_s$ colors
of faint stars in the cluster field are 1.5--2 mag).

\subsection{The Mass Function}

\begin{figure}
\centering
\resizebox{7.6cm}{!}{\includegraphics{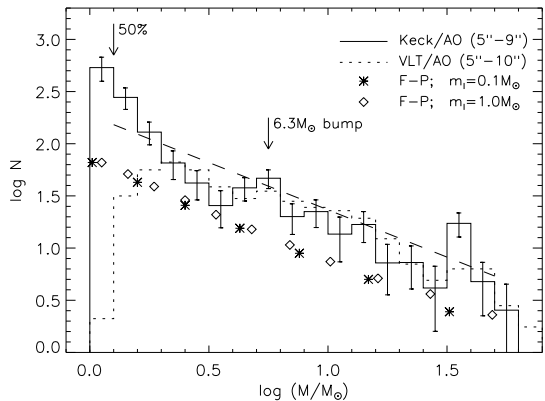}}
\caption{\label{fig:mf}
Background-subtracted mass functions derived from our $K_s$--band luminosity
function for the $5 \arcsec$--$9 \arcsec$ annulus of the cluster field
({\it solid}) and the mass function of Stolte et al. (2005) for the
$5 \arcsec$--$10 \arcsec$ annulus of the
cluster ({\it dotted}).  Error bars and the best-fit power-law relation
($\Gamma = -0.91$)
for $\log M/\msun = 0.1$--1.7 are shown for the former.  The average of
control fields A, B, and C were used as the background for the former, while
the background population was not subtracted for the latter.
Also plotted are the MFs for the same projected annulus from our Fokker-Planck
calculations that fit the observed MF at 2~Myr the best (offset by $-0.5$ dex
for clear presentation).  The {\it asterisks} are for the calculation with
$m_l=0.1 \msun$ and the {\it diamonds} are for $m_l=1 \msun$.  Both
calculations have an initial $\Gamma$ of $-1.1$, and total mass of
$4 \times 10^4 \msun$.  The arrows indicate the locations of the 50~\%
completeness limit and the $6.3 \msun$ bump.}
\end{figure}

We convert the apparent magnitudes into masses using the Geneva models. 
Here, we only use $K_s$--band data as they provide deeper photometry.
Figure~\ref{fig:mf} shows our completeness-corrected MF for the cluster
intermediate region after background subtraction, along with the MF from
Stolte et al. (2005) that is also completeness-corrected, but {\it not} 
background-subtracted; both are 
scaled to an area of $5 \arcsec$--$9 \arcsec$ annulus.  The MF of
Stolte et al. (2005) shows a global turnover below $\log M/\msun \simeq
0.35$ ($M \simeq 2.3 \msun$), which is also their 50~\% completeness limit,
but our cluster MF, whose 50~\% completeness limit is at $\log M/\msun = 0.1$
($M \simeq 1.3 \msun$), still has a significant amount of stars below
$M=2.3 \msun$ and keeps increasing at least down to $M=1 \msun$.  Note that
our MF increases even below our 50~\% completeness limit.

Stolte et al. (2005) claim that there may be a turnover near $\log M/\msun
\simeq 0.8$ ($M \simeq 6.3 \msun$) in their MF and that this might indicate
a global decrease of the background-subtracted MF below that mass.  However,
our data, which have a quarter dex lower completeness limit in mass, indicate
that the MF globally increases down to our completeness limit even after
background subtraction.  Therefore, we presume that the turnover around
$M=6.3 \msun$ claimed by Stolte et al. (2005) is in fact a local bump in
the MF as see in Figure~\ref{fig:mf}.  Eisenhauer et al. (1998) find a
similar bump in their LF of the Galactic starburst template NGC 3603,
and show that such a bump is an indication of the young age ($\lsim 3$~Myr)
using the evolutionary pre-main-sequence tracks by Palla \& Stahler (1993).

When fit to a single power-law relation, our cluster MF gives a power-law
exponent of $\Gamma = -0.91 \pm 0.08$ (the Sapleter slope is $\Gamma = -1.35$)
for the mass range of $\log M/\msun=0.1$--1.7.  This is slightly steeper
than the exponent for massive stars only ($\log M/\msun=0.7$--1.7),
$\Gamma=-0.71 \pm 0.15$.

We have performed several Fokker-Planck calculations (for $m_l = 0.1$ \&
$1 \msun$) and $N$-body simulations (for $m_l =1 \msun$) targeted for the
Arches cluster with initial cluster conditions similar to those used in
Kim, Morris, \& Lee (1999) and Kim et al. (2000), one of which is no initial
mass segretation.  We find that
the $\Gamma$ values for the projected annulus of $5 \arcsec$--$9 \arcsec$
increase by 0.1--0.2 during the lifetime of the Arches cluster,
2--2.5~Myr.\footnote{These calculations consider internal dynamics of the
cluster such as mass segregation and tidal evaporation among others;
see the references for model details.}
Therefore, our results suggest that the IMF of the cluster has
$\Gamma = -1.0 \sim -1.1$.  Figure~\ref{fig:mf} shows two MFs for our annulus
from our Fokker-Planck calculations that best match the observed
MF at 2~Myr.  These two calculations have considerably different $m_l$'s
(0.1 \& $1 \msun$), but result in nearly identical MFs for the annulus.
Thus our estimate for the IMF from the simulations does not sensitively
depend on the choice of $m_l$.

Using the red clump stars in the region $|l| \lsim 2 \arcdeg$ and $0\arcdeg
.5 \lsim |b| \lsim 1\arcdeg$ that were observed with the Infrared Survey
Facility telescope, Nishiyama et al. (2006) newly estimated the near-infrared
extinction law toward the inner Galactic bulge.
When assuming this extinction law, which is derived by averaging for rather
larger region of the GC, is applicable to the stars in our images, we find
that the 50~\% completness limit moves down to $1 \msun$, but there is almost
no change in $\Gamma$ for our annulus ($-0.90 \pm 0.09$).

\acknowledgements
Data presented herein were obtained at the W. M. Keck Observatory, which
is operated as a scientific partnership among the California Institute of
Technology, the University of California, and the National Aeronautics and
Space Administration. The Observatory was made possible by the generous
financial support of the W. M. Keck Foundation.  We appreciate the anonymous
referee for valuable comments, which greatly improved our manuscript.
S.S.K. thanks Myung Gyoon Lee, Mark Morris, Hong Soo Park, and Andrea Stolte
for helpful discussion.  This work was supported by the Astrophysical Research
Center for the Structure and Evolution of the Cosmos (ARCSEC) of Korea Science
and Engineering Foundation through the Science Research Center (SRC) program.
The material in this paper is based upon work supported by NASA under award
No. NNG05-GC37G, through the Long Term Space Astrophysics program.  F.N.
acknowledges PNAYA2003-02785-E and AYA2004-08271-C02-02 grants.


\end{document}